\documentclass[aps,11pt,tightenlines]{revtex4}
\usepackage[utf8]{inputenc}
\usepackage[english]{babel}
\usepackage[titletoc,toc,title]{appendix}
\usepackage{amsmath}
\usepackage{amsfonts}
\usepackage{amssymb}
\usepackage[colorlinks=true,linkcolor=blue,citecolor=red,urlcolor=magenta]{hyperref}
\usepackage{graphicx,float}
\usepackage{enumerate}
\usepackage{subcaption}
\usepackage{color}
\usepackage{mathtools}
\usepackage{hyperref}
\usepackage{comment}

\usepackage{orcidlink}

\definecolor{applegreen}{rgb}{0.0, 0.71, 0.0}

\numberwithin{equation}{section}

\begin{document}

\title{Solutions of an extended Duffing-van der Pol equation with variable coefficients}

 \author{O. Cornejo-P\'erez\orcidlink{0000-0002-1790-7640}}
\email{octavio.cornejo@uaq.mx}
\affiliation{Facultad de Ingenier\'{\i}a, Universidad Aut\'onoma de Quer\'etaro,
Centro Universitario Cerro de las Campanas, 76010 Santiago de Quer\'etaro, Mexico}
\affiliation{Departamento de F\'{\i}sica Te\'orica, At\'omica y Óptica and  Laboratory for Disruptive Interdisciplinary Science (LaDIS), 
  Universidad de Valladolid, 47011 Valladolid, Spain}

\author{P. Albares\orcidlink{0000-0002-2805-9368}}
\email{paz.albares@uva.es}
\affiliation{Departamento de F\'{\i}sica Fundamental, 
  Universidad de Salamanca, 37008 Salamanca, Spain}
\affiliation{Departamento de F\'{\i}sica Te\'orica, At\'omica y Óptica and  Laboratory for Disruptive Interdisciplinary Science (LaDIS), 
  Universidad de Valladolid, 47011 Valladolid, Spain}

\author{J. Negro\orcidlink{0000-0002-0847-6420}}
\email{jnegro@uva.es} 
\affiliation{Departamento de F\'{\i}sica Te\'orica, At\'omica y Óptica and  Instituto de Investigación en Matemáticas (IMUVa), 
  Universidad de Valladolid, 47011 Valladolid, Spain}
	

\begin{abstract}
In this work, exact solutions of the nonlinear cubic-quintic Duffing-van der Pol oscillator with variable coefficients are obtained. Two approaches have been applied, one based on the factorization method combined with the Field Method, and a second one relying on Painlevé analysis. Both procedures allow us to find the same exact solutions to the problem. The Lagrangian formalism for this system is also derived. Moreover, some examples for particular choices of the time-dependent coefficients, and their corresponding general and particular exact solutions are presented.

\bigskip\bigskip

\end{abstract}

\vspace{2pc}

\maketitle

\section{Introduction}

The quest for exact solutions of nonlinear differential equations is an active field of research, as they describe diverse phenomena in physics and engineering systems. In recent decades, there has been a growing interest in studying nonlinear oscillators, with the aim to understand their different properties related to their complex dynamics, and even the rising of chaos. Among them, the Duffing-van der Pol oscillator, presented as a second order ordinary differential equation (ODE), is one of the most widely studied nonlinear systems  \cite{venka1997,lrbook,chandra2005,kudry2018,kimi2009}. This nonlinear ODE has extensive applications in physics, engineering and even biological problems, e.g. a model describing the propagation of voltage pulses along a neuronal axon \cite{venka1997,lrbook}. The force-free Duffing-van der Pol equation is given in the form
\begin{equation}\label{int1}
\ddot{x} - \mu (1-x^2)\dot{x} +\alpha x +\beta x^3 =0, 
\end{equation}
where $\mu (1-x^2)$ is a damping or dissipation term, $\mu>0$ is a constant damping coefficient, and $\alpha$ and $\beta$ are real constant parameters. In \cite{chandra2005}, the general solution of Eq.~\eqref{int1} for certain restrictions in the parameters has been found in implicit form by means of an extension of the Prelle-Singer method. Moreover, in \cite{kudry2018} an explicit special exact solution was obtained by using some effective transformations which linearize Eq. (\ref{int1}). 

Recently, the extended Duffing-van der Pol oscillator, also called cubic-quintic Duffing-van der Pol system or $\phi^6$-van der Pol oscillator, has been studied \cite{siewe2004,ghaleb2021,hu2023,ma2022,zhou2008,getal24}. This extended form of ODE contains a fifth degree monomial term, and the force-free equation is given in the form
\begin{equation}\label{int2}
\ddot{x} - \mu (1-x^2)\dot{x} +\alpha x +\beta x^3 + \delta x^5 =0, 
\end{equation}
where $\delta$ is a real constant parameter. In \cite{siewe2004,ghaleb2021,hu2023}, analytical approximate solutions for some inhomogeneous cases were obtained. Furthermore, extensive numerical analysis for the case of external periodic excitations has been carried out to study its complex dynamics \cite{siewe2004,ghaleb2021,hu2023,ma2022,zhou2008}. In \cite{getal24}, an analytical general solution for Eq. (\ref{int2}) has been derived through the factorization method \cite{berkovich1992,rcp1,rcp2,rcp3,ojlh2005,wang2008}. A variable coefficient generalization for the extended Duffing-van der Pol system including a time-dependent excitation and an external force has been studied numerically in \cite{zhang2022}. 

In this paper, we consider an unforced Duffing-van der
Pol oscillator with cubic and quintic nonlinearities and variable coefficients, referred as DVDP equation in the following, of the form
\begin{equation}\label{int3}
\ddot{x} + \left[ A(t) x^2+ B(t) \right] \dot{x} +  C(t)x +D(t)x^3+E(t)x^5 = 0 , 
\end{equation}
where $A(t)\neq 0, E(t)\neq 0, B(t), C(t), D(t)$ are certain smooth functions of time $t$. We apply two different approaches to search for its general solution: the factorization method in combination with the so-called Field Method \cite{vuj1,vuj2,kov1,kov2}, and Painlev\'e's analysis for integrable nonlinear ODEs \cite{painleve1900,painleve}. Although they may seem completely unrelated, we will show that both procedures impose the same constraints in the coefficients of the nonlinear ODE (for the ODE to be either factorizable or integrable, respectively), therefore leading to the same solution.

The factorization method is a well established technique used to obtain exact solutions of nonlinear ODEs in an algebraic manner. It was widely used for linear ODEs in quantum mechanics after some works of Dirac to solve the spectral problem of the quantum oscillator, and Schrodinger's works on the factorization of the Sturm-Liouville equation. Recently, diverse schemes for nonlinear second and higher order ODEs have been developed, providing many exact solutions of important nonlinear systems \cite{rcp1,rcp2,wang2008,estevez2011,getal22,gonzalez24}. In this work, the advantages arising from the Field Method are used in the framework of the factorization method to obtain the general solution of (\ref{int3}).

In the late XIX century, the primeval works of renowned mathematicians such as Painlevé, Picard, Gambier, Fuchs, etc. \cite{fuchs,gambier,painleve1900,painleve,picard} focused their attention on the classification problem of differential equations according to the types of
singularities of their solutions. Based on these ideas, Painlevé introduced the so-called Painlevé Property for ODEs \cite{painleve1900,painleve}, later extended in a similar fashion to partial differential equations (PDEs) \cite{weiss}, which may act as an integrability criterion. This method has turned out to be a remarkably successful approach to identify integrable families of differential equations, it is connected to other definitions of integrability, and it provides an efficient technique to derive solutions \cite{ince,grammaticos,tabor,estevez1998}. For second order ODEs \cite{painleve,gambier,ince}, there are fifty equations with this property, and then, our goal is to transform Eq.~\eqref{int3} into one these canonical equations. This procedure provides the more general ODE of the form (\ref{int3}) which is integrable in the Painlev\'e sense, leading eventually, to its solutions.


It is worth remarking that the extended equation \eqref{int3} constitutes a broad generalization of the Duffing-van der Pol oscillator that has not been studied previously. The novelty of our research lies in the combination of both approaches to study and successfully provide the exact solution for the integrable version of Eq. \eqref{int3}. The benefits of this synergy have already been highlighted in some previous works \cite{estevez2006,estevez2007}. We propose a novel ansatz in terms of an arbitrary function $f(t)$ to prescribe a noncommutative factorization that allows us to solve the problem and find the associated solutions. From Painlevé’s analysis point of view, the function $f(t)$ arises naturally. In addition to provide the same solution for Eq. \eqref{int3}, this framework also allows us to construct an ideal setting where the derivation of a Lagrangian is immediate through the Jacobi Last Multiplier method \cite{jacobi1,jacobi2,jacobi3,whittaker}.

This paper is organized as follows. In Section \ref{factorization}, a factorization scheme  for nonlinear second order ODEs with variable coefficients, the fundamentals of the Field Method, and an approach combining both methods, are presented. The general and particular solutions for the DVDP equation \eqref{int3} are obtained via the factorization scheme in Section \ref{Duffing}. In Section \ref{sec:PP}, the application of Painlev\'e's integrability techniques is employed so as to find the integrable form of a nonlinear ODE of type \eqref{int3}, and derive its general solution. Section \ref{sec:lagr} is devoted to the Lagrangian formalism for \eqref{int3}. In Section \ref{sec:ex}, three examples of interest for some specific time-dependent coefficients are shown. Finally, in Section \ref{sec:concl} the main conclusions are highlighted.

\section{Factorization of second order nonlinear differential equations and the ``Field Method''}\label{factorization}

Let us consider an even more general ODE than (\ref{int3}), namely the following nonlinear equation with variable coefficients  
\begin{equation}\label{eq1}
\ddot{x} + F(x,t)\dot{x} + G(x,t) = 0 ,
\end{equation}
where $G(x,t)$ and $F(x,t)$ are arbitrary functions of the space and time variables $(x,t)$, respectively. 

Let us  first assume that this equation can be factored in the form
\begin{equation}\label{eq2}
[\mathcal{D}_t-\phi_2(x,t) ] [\mathcal{D}_t-\phi_1(x,t) ]x=0, \qquad  \mathcal{D}_t:=\dot{x}\, \frac{\partial}{\partial x}+ \frac{\partial}{\partial t},
\end{equation}
then, the following conditions \cite{rcp1,rcp2,rcp3} must be met:
\begin{eqnarray}
   \phi_1(x,t)+ \phi_2 (x,t)+ \frac{\partial \phi_1}{\partial x}\, x & =& -  F(x,t), \label{eq3}\\ [1ex]
\phi_1(x,t)\, \phi_2(x,t) -\frac{\partial \phi_1}{\partial t}  & =&  \frac{G(x,t)}{x} .\label{eq3a}
\end{eqnarray}
If now we define $[\mathcal{D}_t-\phi_1(x,t)]x=\Phi(x,t)$, this yields the following coupled system of ODEs for the factored Eq.~(\ref{eq2}),
\begin{eqnarray}
  \dot{x}-\phi_1(x,t)\, x  & =& \Phi(x,t), \label{eq4}\\
\dot{\Phi}(x,t)-\phi_2(x,t)\, \Phi (x,t) & =& 0, \label{eq4a}
\end{eqnarray}
where 
\begin{equation}
\dot{\Phi} \equiv  \mathcal{D}_t \Phi =\frac{\partial\Phi}{\partial t} +\frac{\partial\Phi}{\partial x} \dot{x}.\label{eq4b}
\end{equation}
To find the analytical solution of this system of ODEs, we are going to resort the so-called ``Field Method'' (FM), developed by Vujanovi\'c \citep{vuj1,vuj2}, (see also \citep{kov1,kov2,kovacic2003}). The FM establishes that for a holonomic, nonconservative dynamical system described in the form
\begin{equation}
\dot{x}_j(t) =X_j(t,x_1,...,x_n),\qquad j=1,\dots, n, \label{eq4c}
\end{equation}
one state variable can be chosen to be a field depending on time $t$ and the rest of the variables as
\begin{equation}
x_1 =\Phi (t,x_2,...,x_n). \label{eq4d} 
\end{equation}
Obtaining the total derivative with respect to time of (\ref{eq4d}), and combining it with the last $n-1$ elements of (\ref{eq4c}), we can write the basic field equation as
\begin{equation}
\frac{\partial \Phi}{\partial t} + \sum_{j=2}^{n}\frac{\partial \Phi}{\partial x_j}\,X_j(t,\Phi,x_2,\dots,x_n) - X_1(t,\Phi,x_2,\dots,x_n) = 0, \label{eq4e}
\end{equation}
whose complete solution is of the form
\begin{equation}
x_1=\mathcal{G}(t,x_2,\dots,x_n,\mathcal{C}), \label{eq4f}
\end{equation}
where $\mathcal{C}$ is an arbitrary constant. Then, given (\ref{eq4f}), the solution of Eq. (\ref{eq1}) follows from the system of ODEs (\ref{eq4})-(\ref{eq4a}).

With the aid of \eqref{eq4b}, we start by rewriting the system of Eqs.~(\ref{eq4})-(\ref{eq4a}) as the following quasilinear first order partial differential equation for $\Phi$:
\begin{equation}
\frac{\partial\Phi}{\partial t}+\frac{\partial\Phi}{\partial x} \left( \Phi + x\, \phi_1  \right)-\phi_2\, \Phi=0. \label{eq5}
\end{equation}

As a second step, and following the prescription of previous works on the topic \cite{vuj1,vuj2,kov1,kov2,getal24}, we now look for a complete solution of the above equation in the form
\begin{equation}\label{9}
\Phi(x,t)= \zeta(t)\, x,
\end{equation}
where $\zeta(t)$ is an unknown function of time. Then, by substituting this ansatz in \eqref{eq5}, we get 
\begin{equation}\label{eq6}
\frac{d\zeta(t)}{dt} + \zeta^2(t) - (\phi_2-\phi_1) \zeta(t)=0,
\end{equation}
which necessarily implies the following relation for $\phi_1(x,t)$ and $\phi_2(x,t)$
\begin{equation}
\phi_2 (x,t)= \phi_1 (x,t)+ f(t),
\label{ansf2}
\end{equation}
with $f(t)$ an arbitrary function of time, as long as $\zeta(t)\neq 0$. 

The so-called commutative factorization arises when $f(t)$ is a constant, and then the factorization operators in \eqref{eq2} can be reversed \cite{getal22}. This factorization scheme, when combined with the FM approach, allows to solve important nonlinear ODEs with constant coefficients \cite{getal24}. Nevertheless, in this work, we will consider the general setting where \eqref{ansf2} holds for an arbitrary $f(t)$. The implementation of the general noncommutative factorization and the FM as a technique to deal with differential equations with variable coefficients has not been previously applied.

The relation \eqref{ansf2} allow us now to compute $\phi_1(x,t)$ and $\phi_2(x,t)$ by integrating the system \eqref{eq3}-\eqref{eq3a} in terms of $f(t)$, once the functions $F(x,t)$ and $G(x,t)$ are explicitly given.  By substituting (\ref{ansf2}) into Eq. \eqref{eq3} we get for the factoring functions
\begin{eqnarray}
\phi_{1} (x,t) = \frac{1}{x^2} \left[ h(t) - I(x,t) \right]  - \frac{1}{2}f(t),  \label{eqf1} \\
\phi_{2} (x,t) = \frac{1}{x^2} \left[ h(t) - I(x,t) \right]  + \frac{1}{2}f(t),  \label{eqf2}
\end{eqnarray}
where $I(x,t)= \int^x \tilde{x} F(\tilde{x}, t) d\tilde{x}$, and $h(t)$ is an arbitrary function of time obtained through integration of Eq. \eqref{eq3}. Now, by substituting \eqref{eqf1} and \eqref{eqf2} into Eq. \eqref{eq3a}, we get for the variable coefficients entering the Eq. \eqref{eq1} the relation
\begin{equation}
G(x,t) = \frac{1}{x^3} \left[ h(t) - I(x,t) \right]^2 -\frac{h^\prime (t)}{x} + \frac{1}{2} \left[ f^\prime (t) -\frac{f^2(t)}{2} \right] x + \frac{1}{x}\frac{\partial I(x,t)}{\partial t}. 
\end{equation}
This equation provides the most general relation between $F(x,t)$ and $G(x,t)$ which allows the factorization of Eq. \eqref{eq1} as presented in Eq. \eqref{eq2}.

Furthermore, Eq. \eqref{eq6} therefore provides the following Riccati equation for $\zeta(t)$
\begin{equation} \label{eq7}
\frac{d\zeta(t)}{dt} + \zeta^2(t) -f(t) \zeta(t)=0 ,
\end{equation}
which can be easily integrated as
\begin{equation} \label{zeta}
\zeta(t)= \frac{e^{\int^t dt_1\, f(t_1)}}{k_1 + \int^t  dt_2\left[ e^{\int^{t_2} dt_1\, f(t_1)} \right]}  = \frac{d}{dt} \ln \left(  k_1 + \int^t  dt_2\left[ e^{\int^{t_2} dt_1\, f(t_1)} \right]  \right),
\end{equation} 
in terms of an integration constant $k_1$. 
This last result allows us to find $\Phi(x,t)$ explicitly using \eqref{9}, and consequently the first order ODE (\ref{eq4}) can be written in the form
\begin{equation}\label{eq8}
\dot{x}-\phi_1(x,t)\, x=  \zeta(t) \, x,
\end{equation}
whose integration provides the general solution $x(t)$ for Eq.~(\ref{eq1}) admitting a factored expression in the form proposed in (\ref{eq2}).

We should study separately the case where $\Phi(x,t)=0$, which provides a particular solution for Eq. \eqref{eq1}. Taking the expression for $\phi_1(x,t)$ obtained in the general case, this particular solution is given now by the compatible equation arising from \eqref{eq4}
\begin{equation}
\dot{x}-\phi_1(x,t)\, x= 0. \label{eq8a}
\end{equation}

In the next section we will consider a more specific example of an equation of type \eqref{eq1} in which the functions $F(x,t)$ and $G(x,t)$ have a relatively simple polynomial dependence on the variable $x$, with variable coefficients of time.

\section{Factorization of an extended Duffing-van der Pol equation with variable coefficients}\label{Duffing}

Let us now consider the DVDP equation as in \eqref{int3},
\begin{equation}\label{eq9}
\ddot{x} + \left[ A(t)x^2 +B(t) \right] \dot{x} + C(t)x +D(t)x^3+E(t)x^5 = 0 ,
\end{equation}
where, as already indicated, $A(t)\neq 0,E(t)\neq 0$, and $B(t), C(t), D(t)$ are certain smooth functions of time $t$, which are in principle assumed to be known. The polynomial dependence on $x$ of the functions $G(x,t)$ and $F(x,t)$ is not arbitrary: it has been chosen based on physical applications and interesting mathematical properties \cite{siewe2004,ghaleb2021,hu2023,ma2022,zhou2008}.
Let us proceed with the factorization scheme introduced in Section~\ref{factorization} to solve \eqref{eq9}. 

\subsection{General solution}

Since $\phi_2(x,t)=\phi_1 (x,t)+ f(t)$, equation \eqref{eq9} can be factored according to \eqref{eq2} as follows
\begin{equation}\label{eq11}
[ \mathcal{D}_t - (\phi_1(x,t)+f(t) )] \, [ \mathcal{D}_t -\phi_1(x,t) ]\, x=0,
\end{equation}
and then the following conditions \eqref{eq3}--\eqref{eq3a} are met
\begin{eqnarray}
  2\phi_1(x,t)+ f(t) + \frac{\partial \phi_1(x,t)}{\partial x}\, x  & =&  - \left[ A(t)\, x^2 + B(t) \right], \label{eq12}\\ [1ex]
\phi_1^2(x,t)  +\phi_1(x,t) f(t)  -\frac{\partial \phi_1(x,t)}{\partial t}  & =&  C(t) +D(t)\, x^2+E(t)\, x^4 .\label{eq12a}
\end{eqnarray}
The system above can be straightforwardly integrated, yielding and expression for $\phi_1$ of the form
\begin{equation}\label{eq13}
\phi_1(x,t)= -\frac{A(t)}{4} x^2- \frac{B(t)+f(t)}{2},
\end{equation}
and the following relations for the coefficients
\begin{equation}
E(t)= \frac{A^2(t)}{16},\qquad D(t)=\frac{1}{4}(A(t) B(t) + \dot{A}(t)),\label{eq14b}
\end{equation}
and
\begin{equation}\label{eq14a}
C(t) = \frac{\dot{B}(t)}{2} +\frac{B^2(t)}{4} +\frac{\dot{f}(t)}{2} - \frac{f^2(t)}{4} .
\end{equation}
These results imply that three of the five time-dependent coefficients in \eqref{eq9} actually depend on the other two. These last expressions can be interpreted in two different ways. One is that if in addition to $A(t)$ and $B(t)$ we know the function $f(t)$, then the coefficient $C(t)$ is univocally fixed. Another is that if what is known from the input is the coefficient $C(t)$, the previous expression is a Riccati equation which may determine the possible functions $f(t)$ that allow us to solve the problem by factorization.

In summary, we can say that with the hypotheses used so far, the most general form of the equation \eqref{eq9}  compatible with the factorization proposed is the following 
\begin{equation}\label{eq14d}
\ddot{x} + \left[ A\, x^2 +B \right] \dot{x} + \left(  \frac{\dot{B}}{2} +\frac{B^2}{4} + \frac{\dot{f}}{2} - \frac{f^2}{4}  \right) x  + \frac{1}{4}(AB+\dot{A})x^3 +  \frac{A^2}{16} x^5 = 0 ,
\end{equation}
which admits the decomposition
\begin{equation}\label{eq15}
\left[ \mathcal{D}_t + \frac{A(t)}{4} x^2+\frac{B(t) - f(t)}{2} \right] \left[ \mathcal{D}_t + \frac{A(t)}{4} x^2 + \frac{B(t)+f(t)}{2} \right]x=0.
\end{equation}
If we now assume that solutions of the basic field equation (\ref{eq5}) are factored as in \eqref{9}, then, according to Eq. \eqref{eq8}, we obtain the following compatible first order nonlinear ODE for $x(t)$
\begin{equation}\label{eq16}
\dot{x} +   \frac{R(t)}{2} \, x +\frac{A(t)}{4} x^3=0,
\end{equation} 
where we have defined
\begin{equation}\label{eq1555}
R(t) = B(t)+f(t) - 2 \zeta(t),
\end{equation} 
and $\zeta(t)$ is given by \eqref{zeta} in terms of an arbitrary constant $k_1$.
The general solution of the Bernoulli equation (\ref{eq16}) is also the general solution of (\ref{eq14d}), given in the form
\begin{equation}\label{eq17}
x(t)=  \pm \frac{ \sqrt{2} \, e^{-\frac{1}{2} \int^t dt_1\, R(t_1) }}{\sqrt{ k_2+\int^t dt_2\, A(t_2)  \left[ e^{- \int^{t_2} dt_1\, R(t_1)  } \right]  }} ,
\end{equation}
which depends on two arbitrary constants $k_1$ (which is inside $R(t)$) and $k_2$.

\subsection{Particular solution}

We shall study now the particular solution arising from the choice $\Phi(x,t)=0$ in \eqref{eq4}. This is equivalent to consider
\begin{equation}
\left[ \mathcal{D}_t + \frac{A(t)}{4} x^2 + \frac{B(t)+f(t)}{2} \right]x=0,    
\end{equation}
in \eqref{eq15}, which leads to the following differential equation
\begin{equation}\label{eq16a}
\dot{x} + \frac{B(t)+f(t)}{2}x +\frac{A(t)}{4} x^3=0,
\end{equation} 
whose solution is given by
\begin{equation}\label{eq:32}
x(t)=\pm\frac{\sqrt{2}\,e^{-\frac{1}{2}\int^t dt_1\, \left(B(t_1)+f(t_1)\right)}}{\sqrt{k_3+\int^t dt_2\, A(t_2)  \left[ e^{- \int^{t_2} dt_1\, \left(B(t_1)+f(t_1)\right)  } \right]  }},
\end{equation}
where $k_3$ is the constant of integration.

Particular solutions are obtained by solving the first order compatible ODE $\dot x-\phi_1(x,t)x=0$, where $\phi_1(x,t)$ must be a general solution of \eqref{eq3}. We have considered the same solution for $\phi_1(x,t)$ as in the general case, given in \eqref{eq13}, which follows from the integration of \eqref{eq12}-\eqref{eq12a}. It is worth noticing that this system of PDEs has been solved by imposing $\phi_2=\phi_1+f(t)$, which is a consequence of the ansatz $\Phi=\zeta(t)x$, but no longer holds if $\Phi=0$. Nevertheless, for the sake of simplicity and in order to be able to compare both sets of solutions, we have imposed this ansatz as well in this case. Hence, solution \eqref{eq:32} can be found as the particular case where $\zeta(t)=0$ in the general procedure described previously.

\subsection{Remarks on the noncommutative factorization}

To end this section, and as done in \cite{rcp1}, note that if the order of the brackets is reversed in the factorization of the equation (\ref{eq15}), we arrive at
\begin{equation}\label{eq18}
\left[\mathcal{D}_t + \frac{A(t)}{4} x^2 + \frac{B(t)+f(t)}{2} \right] \left[\mathcal{D}_t + \frac{A(t)}{4} x^2 +\frac{B(t) - f(t)}{2} \right] x=0 ,
\end{equation}
which is a nonlinear ODE of the same type as \eqref{eq14d}
\begin{equation}\label{eq19}
\ddot{x} + \left[ A\,x^2 +B \right] \dot{x} + \left(  \frac{\dot{B}}{2} +\frac{B^2}{4} - \frac{\dot{f}}{2} - \frac{f^2}{4} \right) x 
+\frac{1}{4}(AB+\dot{A})x^3 + \frac{A^2}{16}x^5 = 0,
\end{equation}
where the only difference with \eqref{eq14d} is in the coefficient $C(t)$ through the sign of $\dot f$. Thus, the commutative factorization exclusively arises for $f(t)$ constant. Besides, the sole effect of the noncommutativity of the operators in the factorization is the change $f\to -f$ in the equation, and therefore in its solutions. 


\section{Painlevé's approach}\label{sec:PP}

In the previous section we have analyzed the more general second order ODE of the nonlinear oscillatory type of the form \eqref{eq9} under the factorization technique with reasonable hypothesis. This led us to some restrictions on the values of some of the coefficients of the aforementioned ODE, which becomes \eqref{eq14d}. In the present section we take a different perspective. We use Painlevé's ideas in order to look for the more general ODE of the form \eqref{eq9} that is integrable in the Painlevé sense. This will impose again restrictions over the coefficients, and eventually lead to a solution of the differential equation.

Painlevé, Gambier et al. \cite{painleve,fuchs,gambier} addressed the classification of second order ODEs based on the singularities of their solutions in the complex plane. These authors studied differential equations of the form 
$$\ddot x=\mathcal{F}\left(t,x,\dot{x}\right),$$
where $\mathcal{F}$ is a rational function in $\dot{x}$, algebraic in $x$ and locally analytic in $t$. Painlevé found that there were fifty canonical equations of this form with the property that their critical points are fixed singularities. Forty-four of these equations may be integrated in terms of elementary functions, by quadratures or by linearization. The remaining six equations require the introductions of new trascendental functions, the Painlevé transcendents \cite{clarkson,ince}. These results allowed to introduce the concept of the ``Painlevé Property'' as the following: an ODE has the Painlevé Property if all the movable singularities of its solutions are ordinary poles.    The Painlevé Property can therefore be used as an integrability criterion. If a second order ODE has the Painlevé Property, then it will be integrable in the sense specified before, regardless if we are able to find a explicit solution.

\subsection{Canonical classification}

We aim now at classifying our model \eqref{eq9} as one of those 50 canonical equations established by Painlevé, which have the Painlevé Property by construction. In order to do that, and as prescribed in \cite{painleve,gambier,ince}, we introduce the general scale-like transformation
\begin{equation}
x^2(t)=\lambda^2(t)W(Z), \qquad Z=\varphi(t)    ,
\label{eq:2}
\end{equation}
where from now on $Z$ is the new independent variable, $W(Z)$ is the new dependent variable, and the derivatives with respect to $Z$ are denoted as $'\equiv \frac{d}{dZ}$. The functions $\lambda(t)$ and $\varphi(t)$ are assumed to be smooth and non-zero, and $\dot{\varphi} (t)\neq 0$. By substituting this transformation in \eqref{eq9}, we get
\begin{equation}
\begin{aligned}
W'' & =\frac{{W'}^2}{2W}- \frac{\lambda^2A}{\dot\varphi}\, W\, W' 
-\frac{2\lambda^4E}{{\dot\varphi}^2}W^3-\frac{2\lambda^2}{{\dot\varphi}^2}\left[D+A\frac{\dot\lambda}{\lambda}\right]W^2\\
&\qquad -\frac{1}{\dot\varphi}\left(B+\frac{\ddot\varphi}{\dot\varphi}+\frac{2\dot\lambda}{\lambda}\right)\, W' -\frac{2}{{\dot\varphi}^2}\left[C+B\frac{\dot\lambda}{\lambda}+\frac{\ddot\lambda}{\lambda}\right]W .
\end{aligned}
\label{eq:3}
\end{equation}
If we compare this result with Painlevé's canonical classification \cite{ince}, Eq.~\eqref{eq:3} falls under Type III, and we find that the only possible canonical equation is Eq. XXIV, also known as PXXIV, of the form
\begin{equation}
W''=\frac{m-1}{m}\, \frac{{W'}^2}{W}+q(Z)\, WW'-\frac{m\,  q^2(Z)}{(m+2)^2}\, W^3+\frac{m\, q'(Z)}{m+2}\, W^2,
\label{eq:4}
\end{equation}
where $q(Z)$ is an arbitrary function of $Z$ and $m>1$ is an integer. By comparing the terms in $W'$ and the different powers of $W$ in both \eqref{eq:3} and \eqref{eq:4}, we find that the following conditions need to be met: 
\begin{enumerate}
\item[(a)] The value of the integer $m$ is trivially $m=2$. 

\item[(b)]
The arbitrary function $q(Z)$ turns out to be
\begin{equation}
q(Z)=-\frac{\lambda^2(t)A(t)}{\dot\varphi(t)}   .
\label{eq:6}
\end{equation}
\item[(c)]
Coefficients $E(t)$ and $D(t)$ are identically fixed in terms of $A(t)$, $B(t)$ as 
\begin{equation}
E(t)=\frac{A^2(t)}{16},\qquad D(t)=\frac{1}{4}\left(A(t)B(t)+\dot{A}(t)\right).
\label{eq:5}
\end{equation}
\item[(d)]
The functions $\lambda(t)$ and $\varphi(t)$ are not independent, but they are related through the expression 
\begin{equation}
\label{eq:666}
B+\frac{\ddot\varphi}{\dot\varphi}+\frac{2\dot\lambda}{\lambda}=0.
\end{equation}
Moreover, they must satisfy the following differential equations in terms on $B(t)$ and $C(t)$,
\begin{equation}
\ddot\lambda+B\dot\lambda+C\lambda=0,\qquad\qquad\frac{d}{dt} \left(\frac{{\ddot\varphi}}{\dot\varphi}\right)-\frac 12 \left(\frac{{\ddot\varphi}}{\dot\varphi}\right)^2+
\frac{B^2}{2}+\dot{B}-2C=0
\label{eq:667}
\end{equation}
which are consistent with condition \eqref{eq:666}.
\end{enumerate}

These results indicate that once the coefficients $A(t),\,B(t),\,C(t)$ are known, the most general second order ODE of the form \eqref{eq9} that is integrable in the Painlevé sense is 
\begin{equation}
\ddot{x} + \left[ A\, x^2 +B \right] \dot{x} + C x  + \frac{1}{4}(AB+\dot{A})x^3 +  \frac{A^2}{16} x^5 = 0,
\label{eq:53a}
\end{equation}
where coefficient $B(t)$ and $C(t)$ are no longer independent, since equations \eqref{eq:666} and \eqref{eq:667} must be also satisfied. 

We surprisingly find that this result fully coincides with the one obtained in \eqref{eq14d} using the factorization method. This equivalence straightforwardly holds if we define
\begin{equation}
f(t)=\frac{{\ddot\varphi}}{\dot\varphi},
\label{eq:54a}
\end{equation}
and then, the scale functions $\lambda(t)$ and $\varphi(t)$ are now expressed as
\begin{equation}\label{solphit}
Z=\varphi(t)=\int^t dt_2\, e^{ \int^{t_2} f(t_1)\, dt_1},\qquad \lambda^2(t)=e^{-\int^{t} \left(B(t_1)+f(t_1)\right)\, dt_1}.
\end{equation}
Hence, under these identifications, Eq. \eqref{eq:53a} becomes \eqref{eq14d}, and we conclude that Painlevé integrability for equations of the form \eqref{eq9} implies factorization.

Indeed, this factorization property can be extended to any nonlinear differential equation reducible to the canonical equation PXXIV after a scale transformation \cite{ince}, since it can be easily proven that equation \eqref{eq:4} also admits the commutative factorization
\begin{equation}
\left[ \mathcal{D}_Z -\frac{q(Z)W}{m+2}\right]^2W^{\frac{1}{m}}=0,\qquad \mathcal{D}_Z:=W'\frac{\partial}{\partial W}+\frac{\partial}{\partial Z},   
\label{eq:4.12}
\end{equation}
for any arbitrary function $q(Z)$ and integer $m$. The factorization scheme above is given in terms of $W^{\frac{1}{m}}$, which accounts for the nonlinear contribution in ${W'}^{2}$ in PXXIV \eqref{eq:4} (cf. \cite{gonzalez24}). Such modification arises naturally from the inversion of the scale-like transformation \eqref{eq:2}.

\subsection{Solutions via PXXIV}
According to \cite{painleve,gambier,ince}, Eq. \eqref{eq:4} may be rewritten as the following system of first order ODEs,
\begin{equation}
W'=mWY+\frac{mq(Z)}{m+2}W^2,\qquad\qquad 
Y'=-Y^2,     
\label{eq:53}
\end{equation}
where we have introduced the new dependent variable $Y(Z)$. It is worth noticing that this way of expressing \eqref{eq:4} is not arbitrary, since it is closely related to the factorization for PXXIV. If we define the action of the first operator in \eqref{eq:4.12} as
\begin{equation*}
\left[ \mathcal{D}_Z -\frac{q(Z)W}{m+2}\right]W^{\frac{1}{m}}=YW^{\frac{1}{m}},    
\end{equation*}
then Eq. \eqref{eq:4.12} trivially leads to \eqref{eq:53}. The system \eqref{eq:53} comprises two Riccati equations for $W$ and $Y$, which can be easily integrated by quadratures, yielding the well-known solution for PXXIV \cite{ince} as
\begin{equation}
W(Z)=-\frac{m+2}{m}\frac{\left(C_1Z+C_2\right)^m}{C_3+\int^Z{\left(C_1Z_1+C_2\right)^m\, q(Z_1)\, dZ_1}}.
\label{eq:16}
\end{equation}
where $C_1,C_2,C_3$ are constants of integration, discussed in detail below. 

We may now undo the scale-like transformation \eqref{eq:2} to retrieve the solution $x(t)$ for \eqref{eq9}, yielding
\begin{equation}
x(t)=\pm \frac{\sqrt{2}\,\left(C_1\varphi(t)+C_2\right)\,e^{-\frac{1}{2}\int^tdt_1B(t_1)}}{\,\sqrt{\dot{\varphi}(t)\left(C_3+{\displaystyle \int^t{dt_1\left(C_1\varphi(t_1)+C_2\right)^2\frac{A(t_1)}{\dot\varphi(t_1)}\,e^{-\int^{t_1}{dt_2B(t_2)}}}}\right)}},   
\label{eq:59}
\end{equation}
where $\varphi(t)$ follows from \eqref{eq:667}. 

Solution \eqref{eq:59} may seem to depend on upon three arbitrary constants $C_1,C_2,C_3$. Nevertheless, there are only two relevant constants of integration.  In order to obtain nontrivial solutions for $x(t)$, $C_1$ and $C_2$ cannot vanish simultaneously. Hence, two different cases arise: $C_1\neq 0$ and $C_1=0$. A proper rescaling of the remaining constants and the identification between $f(t)$ and $\varphi(t)$ through \eqref{eq:54a} straightforward yields the solutions obtained through the factorization method \eqref{eq17} and \eqref{eq:32}, respectively.

\section{Lagrangian formalism}\label{sec:lagr}

As it is well known, the formulation of Classical Mechanics in terms of variational principles has proven to be extremely advantageous. It is well established that a standard prescription for the Lagrangian as $\mathcal{L}=T-V$, where $T$ is a quadratic kinetic term and $V$ is a potential function,  mainly works for conservative systems or for specific velocity-dependent forces. Many classical systems exist that do not fall into these categories, such as Liénard-type nonlinear oscillators, but this does not mean that they lack a variational formulation. Nevertheless, the solution to the inverse problem is neither a straightforward nor trivial task. Helmholtz conditions \cite{helmholtz,douglas}, when satisfied, guarantee the existence of a Lagrangian function that gives rise to a given system of ODEs through the Euler-Lagrange equations. Besides, there may exist different (and non-gauge equivalent) Lagrangians for a given system \cite{currie,hojman1,hojman2}. The one-dimensional case (a second order ODE with one generalized coordinate) was first addressed by Darboux \cite{darboux}, and he found that any second order ODE can be derived from a variational problem. Indeed, there exists infinite (typically) non-standard Lagrangians that will yield the desired equation. For this case, Darboux also presented a way to construct the Lagrangian, which reduces to the determination of a function satisfying a differential equation, which turns out to be a Jacobi Last Multiplier (JLM), a concept introduced by Jacobi in the XIX century \cite{jacobi1,jacobi2,jacobi3}. Techniques based on the JLM have proved to retrieve a plethora of remarkably fruitful results when it comes to obtain Lagrangians, either for one-dimensional differential equations \cite{nucci2010,guha2013,carinena2021} or multidimensional systems \cite{nucci2008,nucci2012,carinena2005}.

In this section, we revisit the method of the JLM, focusing on its application to solve the inverse problem. We aim at finding a Lagrangian for equations admitting a factorization scheme as in \eqref{eq14d}, which as we have proven, can also be transformed in an ODE with the Painlevé Property \eqref{eq:4}. Actually, we will start from the Painlevé equation PXXIV, and obtain a Lagrangian by means of the JLM approach for it. Then, the Lagrangian for \eqref{eq14d} arises naturally using the properties of the multiplier.

\subsection{The Jacobi Last Multiplier revisited}

Let us illustrate Jacobi's Last Multiplier method \cite{jacobi1,jacobi2,jacobi3}, first developed as an alternative procedure to derive solutions to mechanical systems that can be reduced to a system of first order differential equations. Let us consider a set of $n$ first order non-autonomous differential equations written in the form
\begin{equation}
\dot{x}_j(t) =X_j(t,x_1,...,x_n),\qquad j=1,\dots, n,  
\label{eq:66}
\end{equation}
where the vector fields $(X_1,\dots,X_n)$ are functions of the $n+1$ variables $(t,x_1,\dots x_n)$. The system above \eqref{eq:66} may be easily rewritten as the Lagrange system
\begin{equation}
dt=\frac{dx_1}{X_1}=\dots=\frac{dx_n}{X_n}.
\label{eq:66b}   
\end{equation}
A function $M=M(t,x_1,\dots x_n)$ is a Jacobi Last Multiplier of the system \eqref{eq:66b} if it satisfies the following differential equation
\begin{equation}
\frac{\partial M}{\partial t}+\sum_{j=1}^n\frac{\partial \left(MX_j\right)}{\partial x_j}=0,
\label{eq:mult1}
\end{equation}
or equivalently, 
\begin{equation}
\frac{d}{dt}\log M+\sum_{j=1}^n\frac{\partial X_j}{\partial x_j}=0,
\label{eq:multi}
\end{equation}
if we take the dynamical system \eqref{eq:66} into account.

Essentially, the JLM formalism states that if $n-1$ first integrals are known, the existence of a last multiplier trivially yields an extra first integral through a quadrature, where the integrating factor depends precisely on the last multiplier \cite{whittaker}. The JLM possesses several properties \cite{jacobi1,jacobi2,jacobi3}, it is closely related to first integrals and Lie symmetries \cite{lie1874,nucci2005}, and it has turned out to be extremely convenient when dealing with Hamiltonian dynamical systems \cite{whittaker}. Of particular interest is the connection between the JLM and the Lagrangian function for second order differential equations.  

Any second order ODE of the form
\begin{equation}
\ddot x=\phi(t,x,\dot x),
\label{eq:76}
\end{equation}
can be alternatively written as the system
\begin{equation}
dt=\frac{dx}{\dot x}=\frac{d\dot x}{\phi(t,x,\dot x)}.  
\label{eq:77}
\end{equation}
Then, it can be proven \cite{whittaker} that a Lagrangian $\mathcal{L}(t,x,\dot x)$ for \eqref{eq:76} can be obtained from the JLM for \eqref{eq:77} as
\begin{equation}
M(t,x,\dot x)=\frac{\partial^2\mathcal{L}}{\partial\dot x^2},   
\label{eq:78}
\end{equation}
since, after this condition, the associated Euler-Lagrange equation 
$$\frac{d}{dt}\left(\frac{\partial\mathcal{L}}{\partial\dot x}\right)-\frac{\partial\mathcal{L}}{\partial x}=0$$
becomes
\begin{equation}
\frac{d}{dt}\log M+\frac{\partial\phi}{\partial \dot x}=0,    
\end{equation}
which is precisely the equation for the Jacobi Last Multiplier of system \eqref{eq:77}.

Hence, a Lagrangian for \eqref{eq:76} can always be obtained from $M(t,x,\dot x)$ \cite{whittaker} as
\begin{equation}
\mathcal{L}(t,x,\dot x)=\int^{\dot x}\left(\int^{\dot x_1} M(t,x,\dot x_2)\, d\dot x_2\right)d\dot x_1+\mathcal{A}(t,x)+\frac{d\mathcal{G}(t,x)}{dt},   
\label{eq:lagr}
\end{equation}
where $\mathcal{G}(t,x)$ is the usual gauge function and $\mathcal{A}(t,x)$ is a function to be determined by imposing that the Euler-Lagrange equation for \eqref{eq:lagr} retrieves \eqref{eq:76}.

Then, the determination of the Lagrangian for a second order ODE reduces to the determination of the JLM for such equation.

\subsection{Non-standard Lagrangians for PXXIV and the extended Duffing-van der Pol equation with variable coefficients}

Let us start rewriting PXXIV \eqref{eq:4} as in \eqref{eq:53},
\begin{eqnarray}
W'=mWY+\frac{mq(Z)}{m+2}W^2,\qquad \qquad 
Y'=-Y^2.\nonumber
\end{eqnarray}
The JLM $M(Z,W,Y)$ for this system arises from \eqref{eq:multi}, and must therefore satisfy
\begin{equation}
\frac{d}{dZ}\log M+mY+\frac{2mq(Z)W}{m+2}-2Y=0.   
\label{eq:mult1a}
\end{equation}
If we assume that $M(Z,W,Y)$ is separable in the following form
\begin{equation}
M(Z,W,Y)=\mu(Z)W^{\alpha}Y^{\beta},    
\end{equation}
where $\alpha,\beta,\mu(Z)$ are parameters to be determined, then, substitution into Eq. \eqref{eq:mult1a}, we get
\begin{equation}
\alpha=-2,\qquad \beta=-m-2,\qquad \frac{d\mu}{dZ}=0.    
\end{equation}
Hence,
\begin{equation}
M(Z,W,Y)=\mathcal{C}W^{-2}Y^{-m-2}   
\end{equation}
constitutes a JLM for the system \eqref{eq:53}, where $\mathcal{C}$ is a trivial scaling constant that plays no role in the dynamics of the system.

It can be easily proven \cite{whittaker} that if we perform a change of variables $(x_1,\dots,x_n)\to (\tilde{x}_1,\dots, \tilde{x}_n)$ in \eqref{eq:66}, the last multiplier $\tilde{M}(t,\tilde{x}_1,\dots,\tilde{x}_n)$ for the transformed system is given by 
\begin{equation}
\tilde{M}(t,\tilde{x}_1,\dots,\tilde{x}_n)=M(t,x_1,\dots,x_n)\,\Delta,  
\label{eq:86}
\end{equation}
where $\Delta$ is the Jacobian of transformation $
\Delta=\frac{\partial(x_1,\dots,x_n)}{\partial(\tilde{x}_1,\dots, \tilde{x}_n)}.$

Then, as done in \cite{nucci2012}, we can straightforwardly compute the Lagrangian for PXXIV by considering the change of variables $(W,Y)\to (W,W')$ in \eqref{eq:53}, giving rise to the following last multiplier
\begin{equation}
\tilde{M}(Z,W,W')=\frac{M(Z,W,Y)}{mW}=\mathcal{C}'W^{m-1}\left(W'-\frac{mq(Z)W^2}{m+2}\right)^{-m-2},  
\end{equation}
where $\mathcal{C}'=m^{m+1}\mathcal{C}$. The multiplier above now satisfies
\begin{equation}
\frac{d}{dZ}\log \tilde{M}+\frac{2(m-1)}{m}\frac{W'}{W}+q(Z)W=0,    
\end{equation}
which is the equation for the JLM associated to \eqref{eq:4}.

Hence, a Lagrangian for PXXIV \eqref{eq:4} can be directly obtained by performing a double integration as in \eqref{eq:lagr}, giving rise to
\begin{equation}
\mathcal{L}(Z,W,W')=W^{m-1}\left(W'-\frac{mq(Z)W^2}{m+2}\right)^{-m}+\frac{d\mathcal{G}(Z,W)}{dZ},
\label{eq:lagrp24}
\end{equation}
where we have identified $\mathcal{C}'=m(m+1)$ for simplicity, $\mathcal{A}(Z,W)=0$, and $\mathcal{G}(Z,W)$ is the gauge function, which can be set to zero without loss of generality. It is easy to check that the Euler-Lagrange equations for \eqref{eq:lagrp24} retrieves \eqref{eq:4}. Lagrangians for Painlevé-type equations via the JLM have been previously studied in literature \cite{ambrosi2009,choudhury2009}, but up to our knowledge so far, the obtention of a Lagrangian for PXXIV is a new result.

Finding now a Lagrangian for the DVDP equation of the form \eqref{eq14d} is immediate by means of the change of variables rule \eqref{eq:86} for the multipliers when applied to the transformation \eqref{eq:2}. Then, a JLM for \eqref{eq14d} is given by
\begin{equation}
\begin{aligned}
N(t,x,\dot x)&=M(Z,W,W')\,\frac{\partial(W,W')}{\partial(x,\dot x)}=4x^2e^{\int^{t}(2B(t_1)+f(t_1)dt_1}M(Z,W,W')\\
&=\frac{3}{2}e^{\int^{t}(2f(t_1)-B(t_1))dt_1}\left[\dot x+\left(\frac{A(t)}{4}x^2+\frac{B(t)+f(t)}{2}\right)x\right]^{-4}, 
\label{eq:multx}
\end{aligned}   
\end{equation}
And finally, a Lagrangian for \eqref{eq14d} arises from \eqref{eq:lagr} as
\begin{equation}
\mathcal{L}(t,x,\dot x)=\frac{1}{4} e^{\int^{t}(2f(t_1)-B(t_1))dt_1}\left[\dot x+\left(\frac{A(t)}{4}x^2+\frac{B(t)+f(t)}{2}\right)x\right]^{-2}. 
\label{eq:lagrx}
\end{equation}
It is worth mentioning the role of the factorization scheme in the derivation of the Lagrangian. A factorization of the form \eqref{eq15} allows us to rewrite the original ODE \eqref{eq14d} as a system of two first order ODEs of Riccati-type,
\begin{equation}
\dot x = yx-\left(\frac{A(t)}{4}x^2+\frac{B(t)+f(t)}{2}\right)x,\qquad\qquad
\dot y=f(t)y-y^2,
\end{equation}
where we have introduced the new dependent variable $y(t)$ as $yx=\left[\mathcal{D}_t-\phi_1(x,t)\right]x$. The system above has a JLM of the form $M(t,x,y)=\mathcal{C}e^{\int^{t}(2f(t_1)-B(t_1))dt_1}x^{-3}y^{-4}$, whose integration provides, after the corresponding change of variables and up a constant factor, a Lagrangian of the form \eqref{eq:lagrx}. So we conclude that the factorization approach provides precisely the ideal setting to compute Lagrangians via the JLM method. Nevertheless, this procedure may not be taken as general, and its application depends on the case of study. 

\section{Examples}\label{sec:ex}

In this section we will present some illustrative examples of interest, either due to their simplicity or their possible applications.

\subsection{Example 1}

Let us consider first Eq. (\ref{eq14d}) in the simplest case of constant coefficients. Let us also assume the following values for the coefficients: $A(t)=A\equiv \text{const.}$, $B(t)=0$, and $C(t)=C\equiv \text{const.}$, which according to Eq. (\ref{eq14a}) implies $f(t)=2\sqrt{C}\textrm{tan}(\sqrt{C}t)$. Then, Eq. (\ref{eq14d}) becomes the nonlinear ODE 
\begin{equation}\label{ex1}
\ddot{x} +  A x^2 \dot{x} + C x + \frac{A^2}{16}x^5 = 0, \, 
\end{equation}
which admits the factorization
\begin{equation}\label{ex2}
\left[ \mathcal{D}_t + \frac{A}{4} x^2 - \sqrt{C}\textrm{tan}(\sqrt{C}t) \right] \left[ \mathcal{D}_t + \frac{A}{4} x^2 + \sqrt{C}\textrm{tan}(\sqrt{C}t) \right]x=0,
\end{equation}
whose general solution is given as follows 
\begin{equation}\label{ex3}
x_{\pm}(t)= \pm \frac{\sqrt{2} \left[ k_1\sqrt{C}\textrm{cos}(\sqrt{C}t) + \textrm{sin}(\sqrt{C}t) \right] }{\sqrt{k_2 C  +\frac{A}{4\sqrt{C}} \left( 2\sqrt{C}\left[ (C k_1^2 +1)t-k_1\textrm{cos}(2\sqrt{C}t) \right] +(C k_1^2 -1)\textrm{sin}(2\sqrt{C}t) \right) } } , 
\end{equation}
where $k_1$ and $k_2$ are integration constants. It is straightforward to see that a bounded general solution can be found in the form
\begin{equation} \label{ex4}
x_{\pm}(t)= \pm \frac{\sqrt{2}}{\sqrt{\frac{1}{2} Ak_1 +\frac{k_2}{k_1^2}\,e^{-\frac{2t}{k_1}}}},
\end{equation}
for $C=-\frac{1}{k_1^2}$, $k_1\neq 0$, which represent a kink-type solution.

One particular solution can also be obtained for Eq.~(\ref{ex1}), and it is provided by Eq. (\ref{eq:32}) as follows 
\begin{equation}  \label{ex5}
x_{\pm}(t)= \pm \frac{\sqrt{2}\textrm{cos}(\sqrt{C}t)}{\sqrt{k_3 + A\left[ \frac{t}{2} + \frac{1}{4\sqrt{C}} \textrm{sin}(2\sqrt{C}t) \right]}} ,
\end{equation}
which is an unbounded solution. 

\begin{figure}[H]
\centering
\includegraphics[width=0.5\columnwidth]{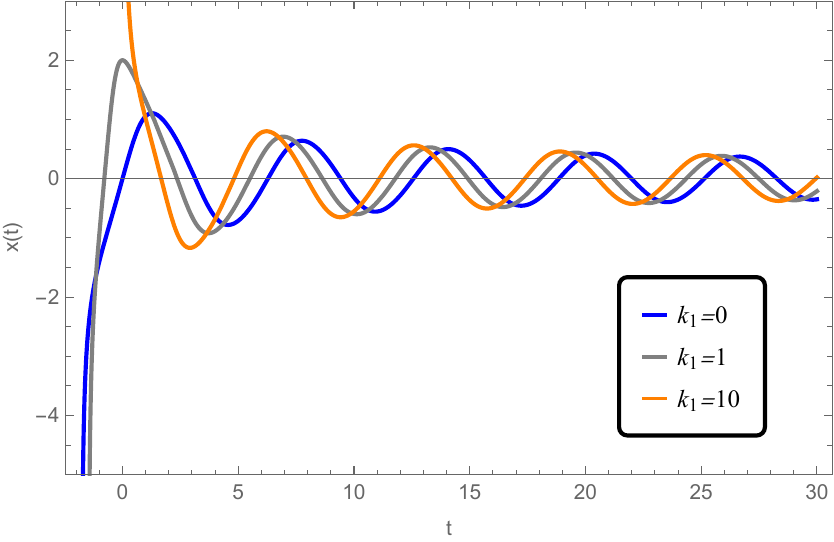}
\includegraphics[width=0.495\columnwidth]{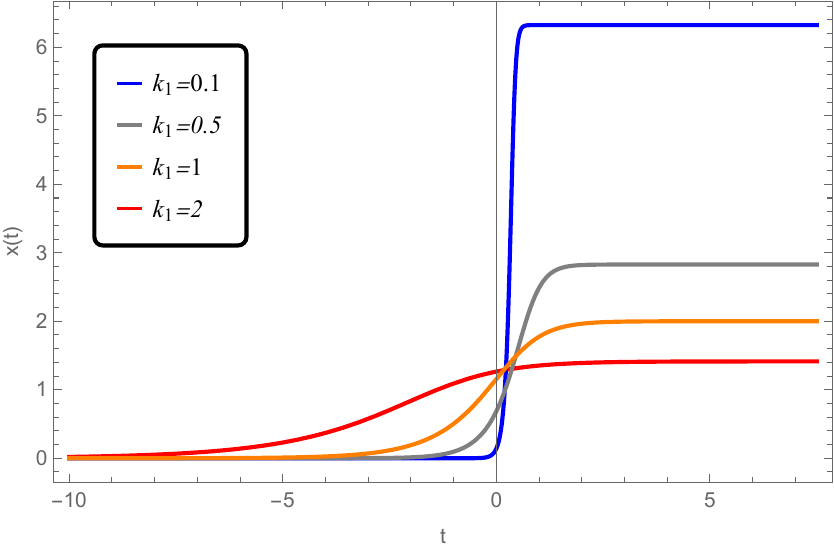}
\includegraphics[width=0.5\columnwidth]{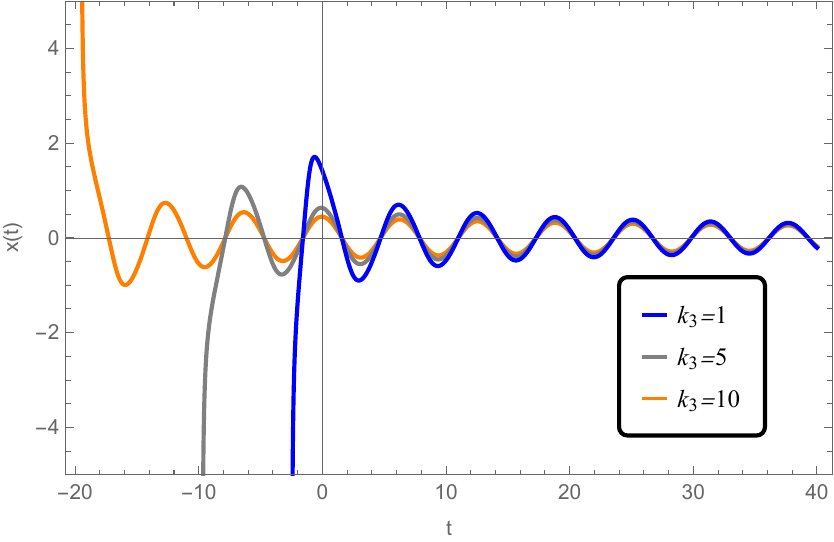}
\caption{General solution $x_{+}(t)$ from (\ref{ex3}) for $A=1$, $C=1$, $k_2=1$, and different values of $k_1$ (upper figure). A similar behavior is obtained by changing the value of $k_2$ and fixing the remaining parameter values. General solution $x_{+}(t)$ from (\ref{ex4}) for $A=1$, $k_2=1$, and different values of $k_1$ as shown in the graphics (middle figure). Particular solution $x_{+}(t)$ in (\ref{ex5}) for $A=1$, $C=1$ and different values of $k_3$ (lower figure).}
\label{fig:1}
\end{figure}

In Figure \ref{fig:1}, plots of the general solutions and the particular solution, for a given set of parameter values, are shown. We can see that the general solution (\ref{ex3}) (upper picture) blows up at a certain time $t=t_0$, whose location can be modified arbitrarily without changing the properties of the dynamics just by carefully selecting the values of $k_1$ and $k_2$ (the effect of varying $k_2$ is a left shift). The system then presents bounded non periodic oscillations with decreasing amplitude, which slowly approaches to zero. The bounded general solution (\ref{ex4}) displays a kink-type pattern which makes a smooth and localized transition between one steady state of null amplitude at $t\to-\infty$ to another steady state of amplitude $\frac{2}{\sqrt{Ak_1}}$ at $t\to\infty$. The particular solution (\ref{ex5}) exhibits a similar dynamics to the one described for the general solution (\ref{ex3}). The sole effect of the parameter $k_3$ is a left shift in the solution. 
Indeed, the particular solution (\ref{ex5}) can be understood as the limit curve of (\ref{ex3}) when $k_1\to\infty$, and then $k_2$ plays essentially the same role as $k_3$.

It is worth remarking that the values chosen for the different parameters involved must guarantee that the final solution to \eqref{eq9}, either general or particular, is real and continuous in the domain of interest. This consideration also applies in the forthcoming examples.

According to \eqref{eq:lagrx}, the Lagrangian for this case explicitly reads
\begin{equation}
\mathcal{L}(t,x,\dot x)=\frac{1}{4}\sec^4(\sqrt{C}t)\left[\dot x+\frac{A}{4}x^3+\sqrt{C}x\tan(\sqrt{C}t)\right]^{-2}.    
\end{equation}

\subsection{Example 2}

Let us consider now the following time-dependent coefficients with a linear polynomial form $A(t)=a_1 t +a_2$, $B(t)=b_1 t$, $b_1\neq 0$, and the simplest case where $f(t)=0$. Then, the nonlinear ODE \eqref{eq14d} is
\begin{equation}\label{ex6}
\ddot{x} + \left[(a_1 t +a_2) x^2 + b_1 t \right] \dot{x} + \frac{b_1}{2} \left( 1 + \frac{b_1 t^2}{2}  \right) x 
+\frac{1}{4} \left[ (a_1t+a_2)b_1t + a_1 \right] x^3 + \frac{1}{16} (a_1t+a_2)^2 x^5 = 0  ,
\end{equation}
 which admits the (commutative) factorization 
\begin{equation}\label{ex7}
\left[ \mathcal{D}_t +  \frac{(a_1t+a_2)}{4} x^2 + \frac{b_1 t}{2} \right]^2 x=0.
\end{equation}
 The general solution of Eq. (\ref{ex6}) is given as follows 
\begin{equation}\label{ex8}
x_{\pm}(t) = \pm \frac{2b_1 \left(t+ k_1 \right) }{ 
\sqrt{2k_2b_1^2e^{\frac{b_1t^2}{2}}+\tilde{c} e^{\frac{b_1t^2}{2}}\operatorname{erf}\left( \frac{\sqrt{b_1}t}{\sqrt{2}} \right) -2a_2b_1(t+2k_1) -2a_1\left[2+b_1(t+k_1)^2 \right] }} ,
\end{equation}
 where $\tilde{c}= \sqrt{2\pi b_1} \left( a_2 +2a_1k_1 +a_2b_1k_1^2 \right)$, $k_1$ and $k_2$ are integration constants. In addition, according to Eq. (\ref{eq:32}), one particular solution is obtained as follows
\begin{equation}\label{ex9}
x_{\pm}(t) = \pm \frac{ 2\sqrt{b_1} e^{-\frac{b_1t^2}{4}} }{ 
\sqrt{2b_1 k_3 + a_2\sqrt{2\pi b_1}\operatorname{erf}\left( \frac{\sqrt{b_1}t}{\sqrt{2}} \right) -2a_1e^{-\frac{b_1t^2}{2}}}} ,
\end{equation}    
 
\begin{figure}[H]
\centering
\includegraphics[width=0.5\columnwidth]{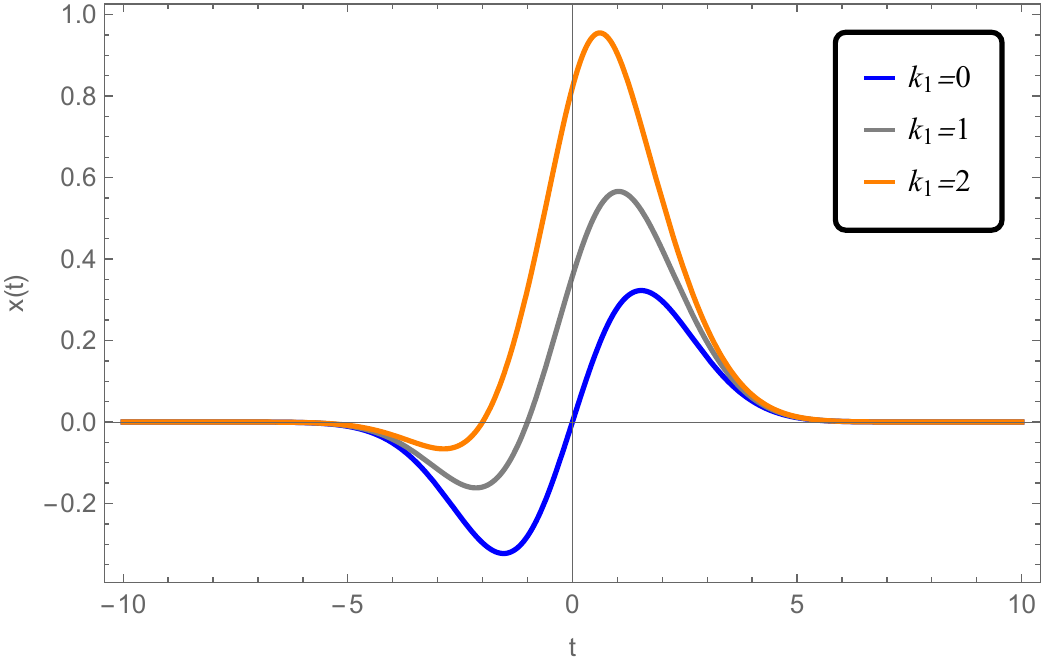}
\includegraphics[width=0.5\columnwidth]{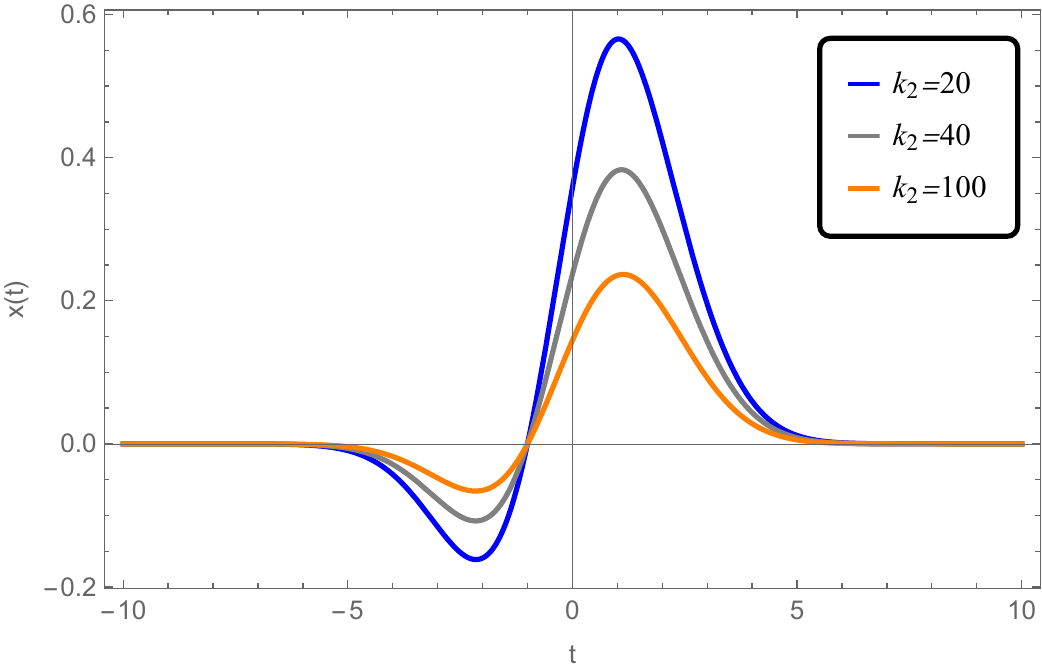}
\includegraphics[width=0.5\columnwidth]{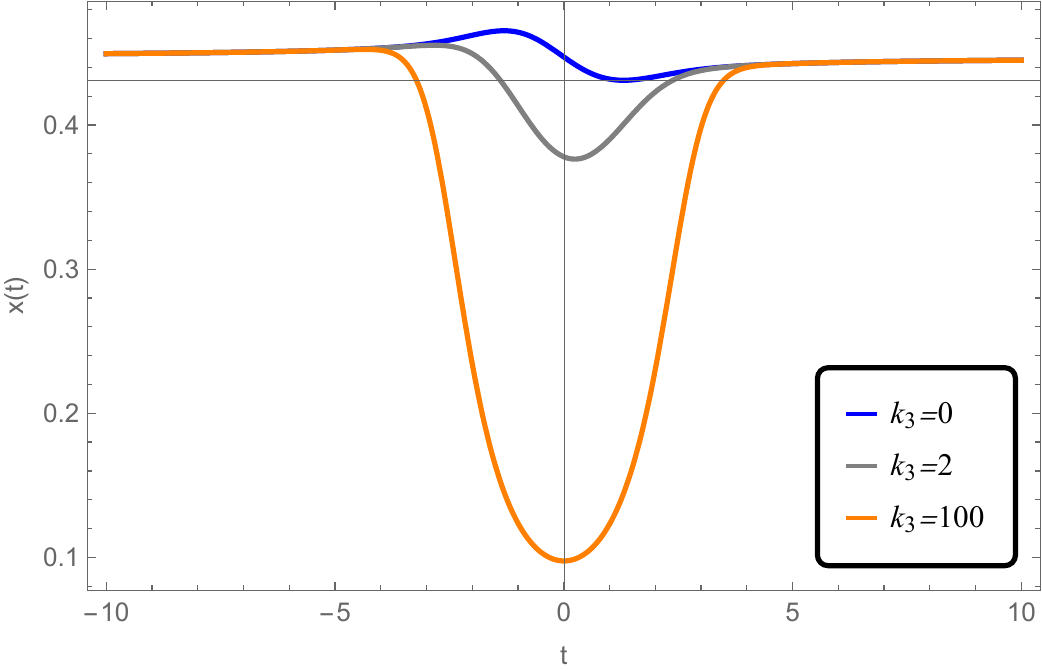}
\caption{General solution $x_{+}(t)$ from (\ref{ex8}) for $a_1=1$, $a_2=0$, $b_1=0.8$, $k_2=20$, and different values of $k_1$ (upper figure). General solution $x_{+}(t)$ from (\ref{ex8}) for $a_1=1$, $a_2=0$, $b_1=0.8$, $k_1=1$, and different values of $k_2$ (middle figure). Particular solution (\ref{ex9}) for $a_1=10$, $a_2=1$, $b_1=-1$ and different values of $k_3$.}
\label{fig:2}
\end{figure}

In Figure \ref{fig:2}, plots of the general solution (\ref{ex8}) and particular solution (\ref{ex9}) are displayed. For a precise balance in the parameters, it is possible to obtain solutions that are bounded and exhibit a localized behaviour in time, as illustrated in the pictures. In the general solution case (\ref{ex8}) (upper and middle figures), the system oscillates once near the origin and then decays to zero at $\left|t\right|\to\infty$. The particular solution \eqref{ex9} however displays a slightly different behaviour (lower figure). Starting from the constant value $\sqrt{-\frac{2b_1}{a_1}}$, the system performs one oscillation, whose amplitude and width increase as $k_3$ becomes larger, reaching the same asymptotic value as $t\to\infty$.  

The Lagrangian for Eq. \eqref{ex6} now reads
\begin{equation}
\mathcal{L}(t,x,\dot x)=\frac{1}{4}e^{-\frac{b_1}{2}t^2}\left[\dot x+\frac{1}{4}\left(a_1x^2+2b_1\right)tx+\frac{a_2}{4}x^3\right]^{-2}.    
\end{equation}

\subsection{Example 3}

Another example arises if we assume the following forms for $A(t)$, $B(t)$ and $f(t)$:
\begin{equation}\label{eq2333}
A(t)= - B(t)= \mu (t) = \frac{1}{2}\left[ 1+ \tanh t  \right] >0, \quad  f(t)=f\equiv \text{const.}, f \in R - \{0\}.
\end{equation}
The nonlinear second order ODE \eqref{eq14d} now becomes 
\begin{equation}\label{eq23}
\ddot{x} -  \mu (t) \left[1- x^2 \right] \dot{x} + \frac{1}{4} \left(  \mu^2(t) - 2 \dot{\mu}(t) - f^2 \right) x 
+\frac{1}{4} \left( \dot{\mu}(t) - \mu^2 (t)\right) x^3 + \frac{\mu^2(t)}{16}x^5 = 0  ,
\end{equation}
which admits the following factorization 
\begin{equation} \label{eq23a}
\left[ \mathcal{D}_t + \frac{\mu(t)}{4} x^2 - \frac{\mu(t) +f}{2} \right] \left[ \mathcal{D}_t + \frac{\mu(t)}{4} x^2 - \frac{\mu(t)-f}{2} \right]x=0 .
\end{equation}
The general solution of Eq. (\ref{eq23}) is given in analytic form by 
\begin{equation}\label{eq23b}
x_{\pm}(t) = \pm \frac{\sqrt{2}\left(1+e^{2t}\right)^{\frac{1}{4}}\left(k_1 +\frac{1}{f} e^{ft} \right) e^{-\frac{ft}{2
}}}{ \sqrt{\sqrt{2}k_2 + \int^t dt_1 \left(k_1 +\frac{1}{f} e^{ft_1} \right)^2 e^{(2-f)t_1}\left(1+e^{2t_1}\right)^{-\frac{1}{2}}}} ,
\end{equation}
with $k_1$ and $k_2$ as integration constants. The particular solution is 
given as follows 
\begin{equation}\label{eq23c}
x_{\pm}(t) = \pm \frac{\sqrt{2}\left(1+e^{2t}\right)^{\frac{1}{4}}e^{-\frac{ft}{2
}}}{ \sqrt{\sqrt{2}k_3 + \int^t dt_1 e^{(2-f)t_1}\left(1+e^{2t_1}\right)^{-\frac{1}{2}}}}.
\end{equation}

Both solutions can be integrated, for the appropriate ranges of the parameters, as a combination of hypergeometric and elementary functions. 

\begin{figure}[H]
\centering
\includegraphics[width=0.48\columnwidth]{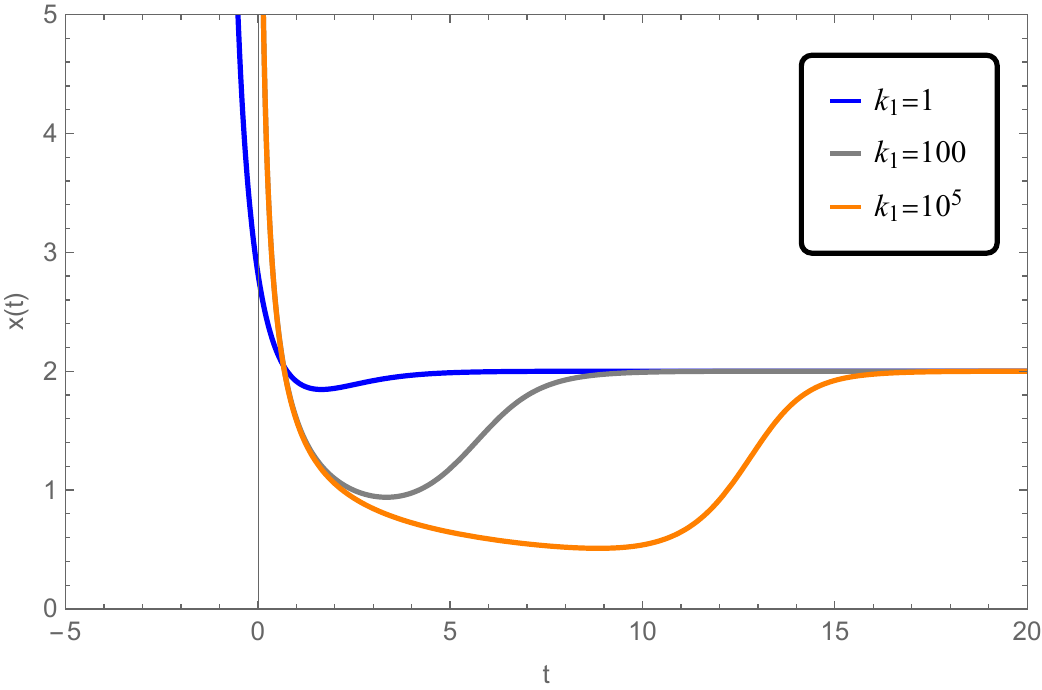}
\includegraphics[width=0.48\columnwidth]{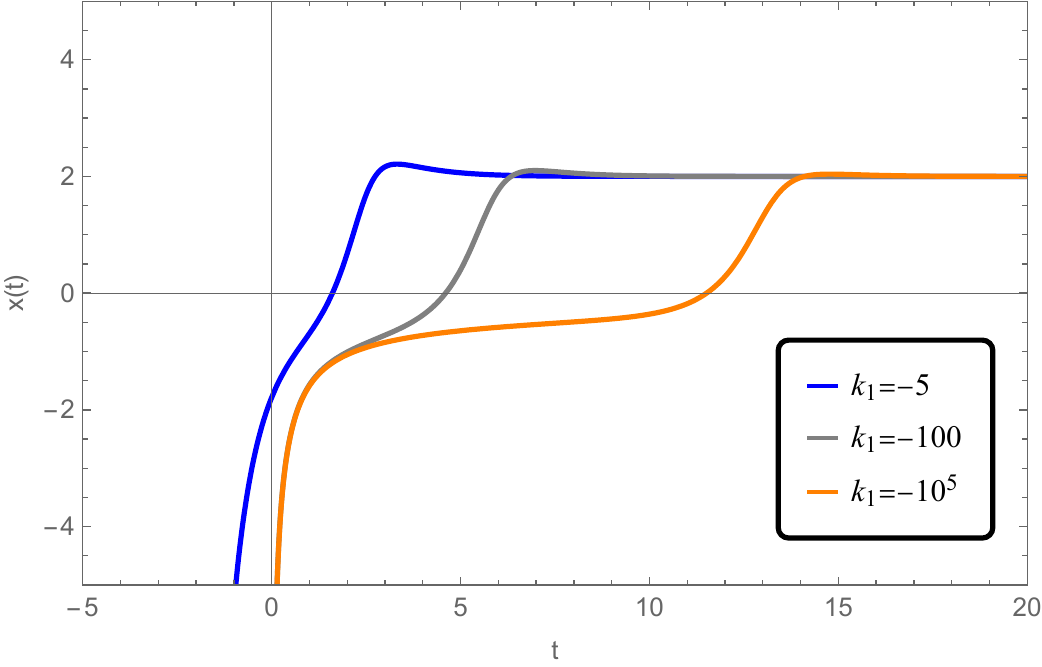}
\includegraphics[width=0.48\columnwidth]{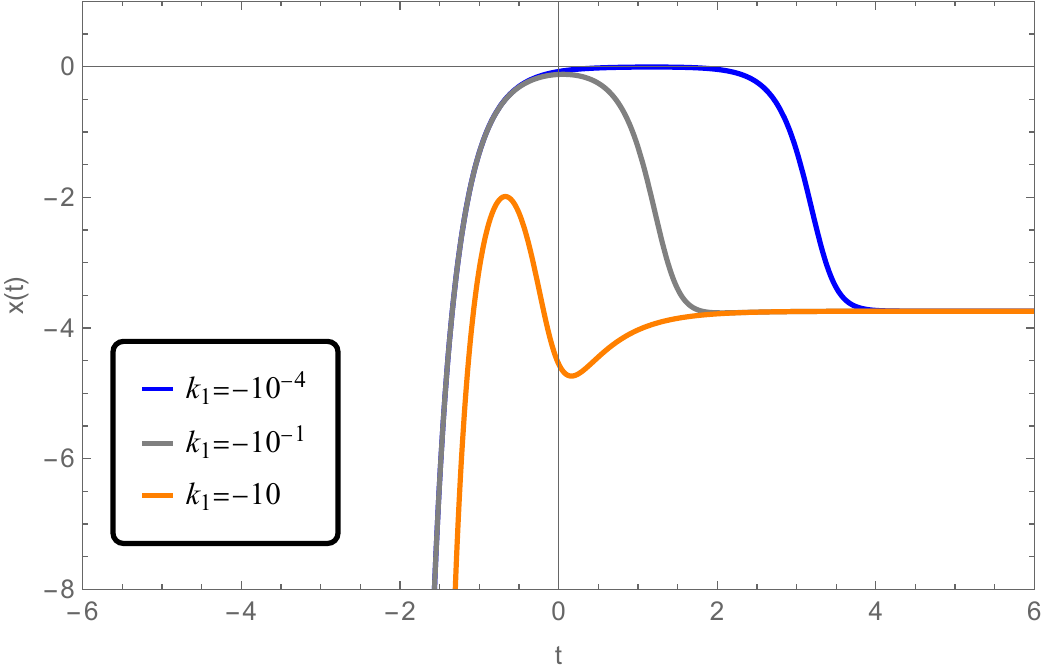}
\includegraphics[width=0.48\columnwidth]{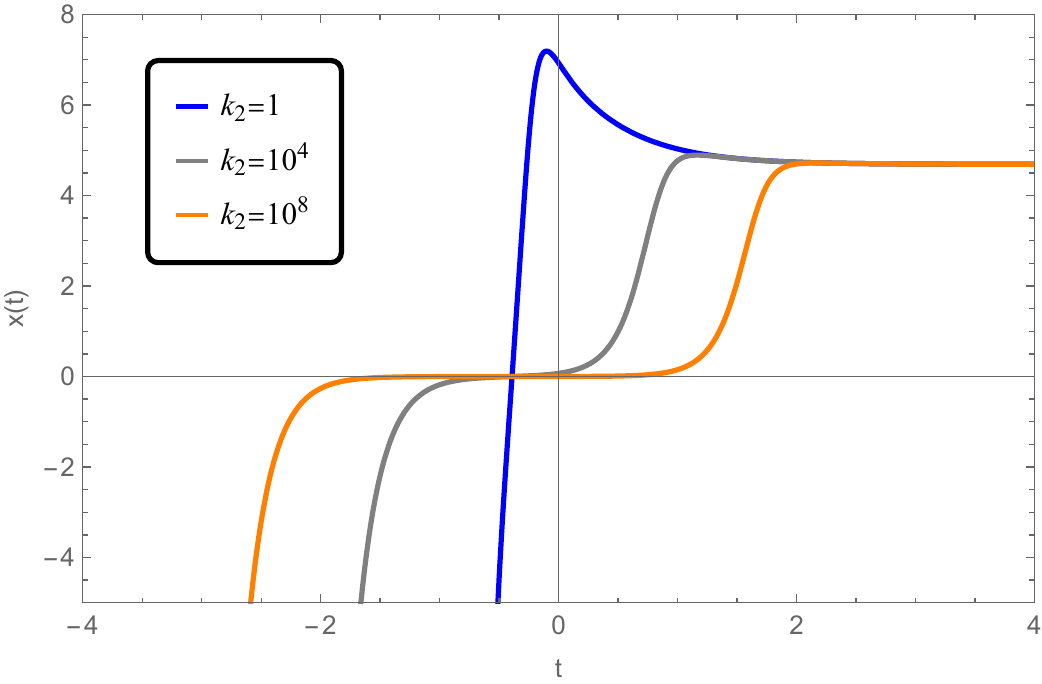}
\caption{General solution $x_{+}(t)$ from (\ref{eq23b}) for different choices of the parameters: $f=1$, $k_2=1$, and different values of $k_1>0$ (upper left figure); $f=1$, $k_2=10$, and different values of $k_1<0$ (upper right figure); $f=-6$, $k_2=10$, varying $k_1<0$ (lower left figure); $f=-10$, $k_1=5$, and increasing values of $k_2>0$ (lower right figure).}
\label{fig:3}
\end{figure}

Figure \ref{fig:3} displays some interesting dynamics for the general solution (\ref{eq23b}). Overall, a divergence is always present, whilst the system tends to the asymptotic value $\pm\sqrt{2(1+\left|f\right|)}$ at $t\to\infty$. Nevertheless, diverse behaviours arise for different choices of the parameters involved. For $f>0$, the general solution may show a semi-infinite well type shape (left upper figure) for $k_1>0$, or a barrier type solution with an intermediate step (right upper figure) if $k_1<0$. For $f<0$, a semi-infinite barrier can emerge if $k_1<0$ (lower left figure), while for a fixed $k_1>0$, an intermediate flat plateau may appear at $x=0$ for large values of $k_2$ (lower right figure).

\begin{figure}[H]
\centering
\includegraphics[width=0.5\columnwidth]{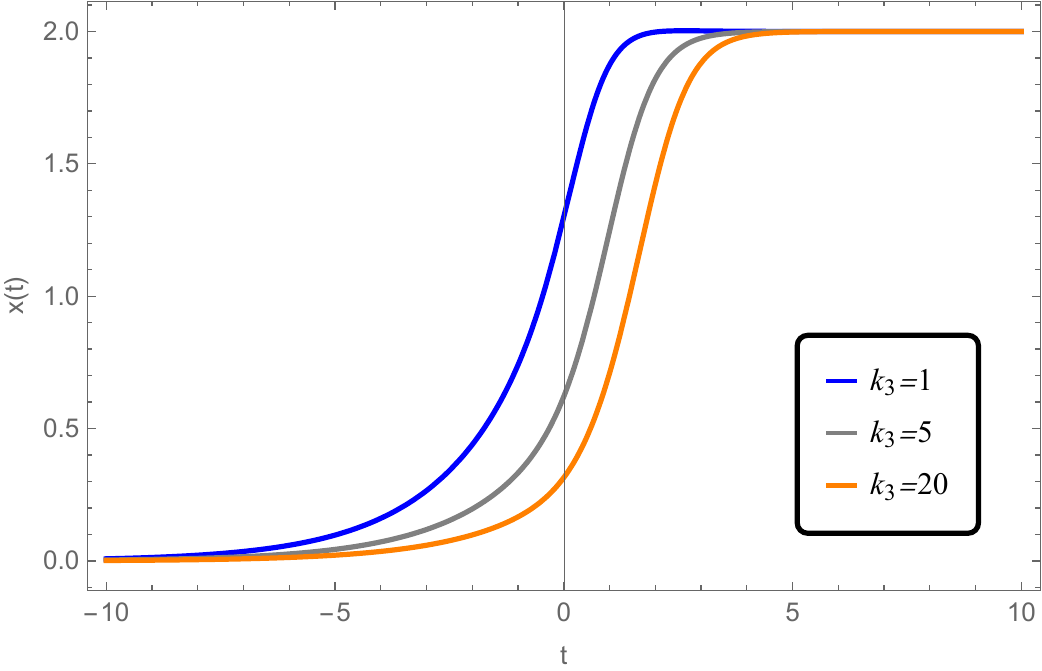}
\includegraphics[width=0.5\columnwidth]{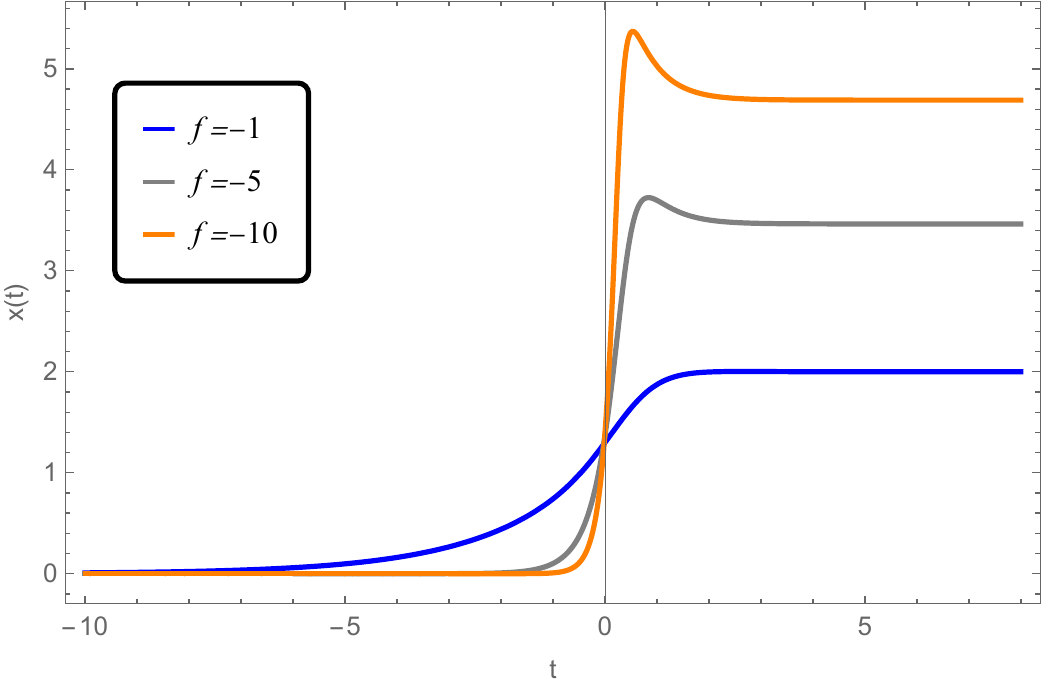}
\caption{Particular solution (\ref{eq23c}) for $f=-1$ when increasing $k_3$ (upper figure), and for $k_3=1$ while varying $f$ (lower figure).}
\label{fig:4}
\end{figure}

In Figure \ref{fig:4}, the behaviour of particular solution (\ref{eq23c}) is analyzed. This solution displays a kink-type profile for $f<0$, where the system transitions between two stationary states of asymptotic null amplitude at $t\to-\infty$ and $\sqrt{2(1-f)}$ at $t\to\infty$. Such behaviour is evinced in two cases: when changing $k_3$ for a fixed $f$ (upper figure), and when $f$ varies for a given $k_3$ (lower figure). 

Finally, the Lagrangian for this case turns out to be
\begin{equation}
\mathcal{L}(t,x,\dot x)=\frac{1}{4\sqrt{2}}e^{2ft}\sqrt{1+e^{2t}}\left[\dot x+\left(\frac{\mu(t)}{4}x^2+\frac{f-\mu(t)}{2}\right)x\right]^{-2}.
\end{equation}

\section{Conclusion}\label{sec:concl}

In this work, we have obtained exact solutions of the extended Duffing-van der Pol oscillator with variable coefficients, given in the integrable form (\ref{eq14d}). The fact that Eq.~\eqref{eq14d} admits a noncommutative factorization imposes some restrictions for the time-dependent coefficients: $C(t)$, $D(t)$, and $E(t)$ depend on $A(t)$, $B(t)$, and the arbitrary function $f(t)$. Then, the factorization scheme, together with the FM, provides the general solution to the problem as well as a particular solution, which is obtained by solving the corresponding compatible first order ODE. The combination between these two techniques as a way to find solutions to nonlinear ODEs with variable coefficients is a novel approach.

The relations found for the coefficients exactly coincide with the ones obtained through Painlevé analysis. These restrictions precisely arise from the imposition that the general DVDP equation \eqref{eq9} is transformed into PXXIV \eqref{eq:4} after a scale-like transformation. PXXIV can be straightforwardly integrated, yielding the solution for \eqref{eq14d}. This solution is proved to fully coincide with the one derived through the factorization method.

Another indication of the powerful convergence of both approaches is the derivation of the Lagrangian. The factorization scheme provides the optimal setting to compute a multiplier and a Lagrangian for either PXXIV and the DVDP oscillator. Since both equations are connected through a scale transformation, the multipliers and Lagrangians for each case are related to each other. It is worth noticing that the combination of the factorization method with the Jacobi Last Multiplier as a way to find Lagrangians is a new approach, as well as the results derived from it.

In order to show the rich dynamical behaviour of the force-free nonlinear DVDP oscillator, three illustrative examples are presented. Plots of the general and particular solutions for all three cases have been displayed and deeply analyzed. We have also checked that the numerical integration of the DVDP Eq. in each case successfully agrees with the analytical solutions obtained throughout this research.

\section{Acknowledgements}
This research was supported through the QCAYLE project,
funded by the European Union-Next Generation UE/MICIU/Plan de Recuperacion, Transformacion y Resiliencia/Junta de Castilla y Leon (PRTRC17.11), and also by RED2022-134301-T and PID2020-113406GB-I00, both financed by MICIU/AEI/10.13039/501100011033. P. Albares acknowledges support from a postdoctoral fellowship “Margarita Salas para la formación de jóvenes doctores”, funded by the European Union-NextGenerationEU and the Spanish Ministerio de Universidades.
\bigskip

\color{black}

\end{document}